\documentclass[conference]{IEEEtran}
\ifCLASSINFOpdf
   \usepackage[pdftex]{graphicx}
   \DeclareGraphicsExtensions{.pdf,.jpeg,.png}
\else
\fi

\usepackage{amsmath}
\usepackage{algorithm}
\usepackage{algpseudocode}
\usepackage{cite}
\usepackage{multicol}
\usepackage{pifont}
\usepackage{siunitx}
\usepackage{subfig}
\usepackage{atbegshi,picture} 

\makeatletter
\def\BState{\State\hskip-\ALG@thistlm}
\makeatother

\AtBeginShipout{\AtBeginShipoutUpperLeft{
  \put(\dimexpr\paperwidth-1cm\relax,-1.5cm){\makebox[0pt][r]{\framebox{This paper was presented in IEEE MASS REU workshop, 2019, Monterrey, CA}}}
}}

\begin{document}

\title{Server-side Fingerprint-Based Indoor Localization Using Encrypted Sorting}

\author{
\IEEEauthorblockN{Andrew Quijano}
\IEEEauthorblockA{
Dept.  of Computer Science\\
Columbia University\\
New York, NY 10025\\
Email: afq2101@columbia.edu}
\and
\IEEEauthorblockN{Kemal Akkaya}
\IEEEauthorblockA{
Dept. of Electrical \& Computer Engineering\\
Florida International University\\
Miami, Florida 33174\\
Email: kakkaya@fiu.edu}
}

\maketitle
\begin{abstract}
GPS signals, the main origin of navigation, are not functional in indoor environments. Therefore, Wi-Fi access points have started to be increasingly used for localization and tracking inside the buildings by relying on fingerprint-based approach. However, with these types of approaches, several concerns regarding the privacy of the users have arisen. Malicious individuals can determine a clients daily habits and activities by simply analyzing their wireless signals. 
While there are already efforts to incorporate privacy to the existing fingerprint-based approaches, they are limited to the characteristics of the homomorphic cryptographic schemes they employed. In this paper, we propose to enhance the performance of these approaches by exploiting another homomorphic algorithm, namely DGK, with its unique encrypted sorting capability and thus pushing most of the computations to the server side. We developed an Android app and tested our system within a Columbia University dormitory. Compared to existing systems, the results indicated that more power savings can be achieved at the client side and DGK can be a viable option with more powerful server computation capabilities. 
\end{abstract}


\IEEEpeerreviewmaketitle
\section{Introduction}
While GPS has been revolutionary in its ability to easily locate a person outdoors accurately and provide directions to travel to their desired destination, it is not as useful to localize users indoors as it was not designed for that niche. There are various technologies that can overcome the shortcomings of GPS for indoor localization such as taking advantage of Wi-Fi Access Points (APs), sensors and RFID devices \cite{hasan2018overview}. One of such promising technologies is based on fingerprinting concept where  Wi-Fi signal strengths (i.e., Received Signal Strength (RSS)) on a floor of a building are pre-collected and shortest distances to those RSS values are computed to determine a user's location \cite{basri2016survey}. The pre-collection phase is called training and required before the localization can be done.  

While there has been an extensive amount of research conducted to improve the performance and granularity of indoor localization, user privacy is usually not the first priority. Nevertheless, there is a growing concern among users that either malicious individuals or large enterprises such as Google use indoor localization systems to compromise their privacy \cite{sadhu2017collabloc}. Since smart phones are ubiquitous, it is conceivable that a malicious individual can collect their victim's MAC Address and data such as RSS to track the daily habits of their target when fingerprint-based approaches are used. 

To address this growing concern, a number of solutions have been proposed in the literature \cite{konstantinidis2015privacy}. The main objective is to either hide the identity of the user or prevent the AP/server from easily viewing the user's data. In case of fingerprint-based solutions, the approach was to hide the RSS data of user devices from the server by employing cryptographic techniques such as homomorphic encryption which enables computation on the encrypted data. In these approaches, the RSS values are encrypted before they are sent to a cloud server and the computations at the server is done on the encrypted data. 


Recent approaches, utilized partially homomorphic systems to ensure privacy \cite{TONYALI2018547, saputro2014preserving, ORIGINAL}. Typically, Paillier-based systems \cite{paillier1999public} are used to compute the distances and send them back to the user device in the encrypted form where they were decrypted to find out the minimum. This is needed because even though the distances can be computed and obtained in the encrypted form, they cannot be compared to each other or sorted in a list as Paillier cannot provide such a feature. 


In this paper, we propose to extend the Paillier-based approaches by designing an alternative system where the computations for sorting the encrypted values are outsourced to the server rather than the client. This stems from the motivation that the server would have lots of computational power and through such outsourcing the client device can save more energy. To this end, we adopted the homomorphic \cite{veugen2018correction} algorithm which has the capability to compare two encrypted values named after the authors Damg{\aa}rd, Geisler and Kr{\o}igaard (DGK). Basically, we introduce an encrypted sorting mechanism at the server and relay only the top result(s) (i.e., $k$ minimum values) to the client.  


We implemented an Android app that uses both systems which can localize users in the third floor of the Broadway dormitory at Columbia University. 
For comparison, we implemented a Java package containing the DGK and Paillier homomorphic encryption systems both at the client and server sides and studied the performance trade-offs. The results indicated that while DGK is faster than Paillier in both a client or server localization approach. In addition, our hypothesis of outsourcing the encrypted comparison in our server side localization approach increased computation time but we used less battery after repeated testing. 

The structure of this paper is as follows. In the next section, we summarize the related work. Section III describes the background which include fingerprint-based localization, homomorphic encryption and its applications in our system. Section IV describes how to initialize the secure fingerprint database and localization approach. Section V provides a comprehensive performance evaluation of both the client and server-based indoor localization system. Finally, Section VI concludes the paper.

\section{Related Works}
There has been a number of fingerprint-based approaches in the literature. For instance, Ahmad et al. follow an approach in the sense that all localization data is to be stored on the individual device rather than a centralized remote server \cite{ahmad2016passive}. Their model uses a Wi-Fi fingerprint database to localize a user and their system can dynamically update itself as the phone is consistently collecting data. It has been proven to be as accurate as traditional centralized fingerprint databases as it also passive collects beacon-stuffing information from APs as well. Beacon stuffing releases information to the phone such as Region ID, Location Number, AP ID, and AP fingerprint that is recorded at its location  \cite{ahmad2016passive}. 
This system provides essentially perfect security to the user, assuming the phone is not compromised. Yet, there is a danger to the network administrator as the APs leak valuable information that could be useful of a malicious individual's reconnaissance phase in order to attempt to hack the network. Therefore, it is likely that network administrators would probably disable this information being leaked.

Zhu et al. have created an indoor localization system using a differentially private algorithm designed to both protect the fingerprint database and the user's data \cite{zhu2017wifi}. A common definition of differentially private is when looking at the output of the algorithm, a third party cannot distinguish any uniquely identifiable data was included in the original dataset or not. Therefore, they used the exponential mechanism on their dataset to conceal the identifiable attributes. Finally they used the J48 Decision tree classifier and showed that their system is still about 90\% accurate in localizing via Wi-Fi fingerprint data. Zhu et al. mention that they tested their indoor location system only on synthetic data. Therefore, there is a possibility that it may have a reduced accuracy when it is tested indoors. Also, there is no discussion on the computational overhead for the process of creating a differentially private system when it comes mobile devices to localize with the assistance of a fingerprint database server. 
While mobile phones have improved dramatically in their computational power, they will consistently have less computational resources or battery lifetime than a laptop.

In this work, we improve the work in \cite{ORIGINAL}. In this work, the Paillier system has been used to perform encrypted computations and send them back to the client. We will revise this approach to utilize DGK \cite{veugen2018correction} which is a homomorphic system with encrypted computing and comparing capabilities and thus push the burden to the server side. 

\section{Background}
\subsection{Fingerprint-based Localization}
\begin{figure}[htb]
\setlength{\abovecaptionskip}{0cm}
\setlength{\belowcaptionskip}{0cm}
    \begin{center}
    \includegraphics[width=\linewidth, height=1.5in]{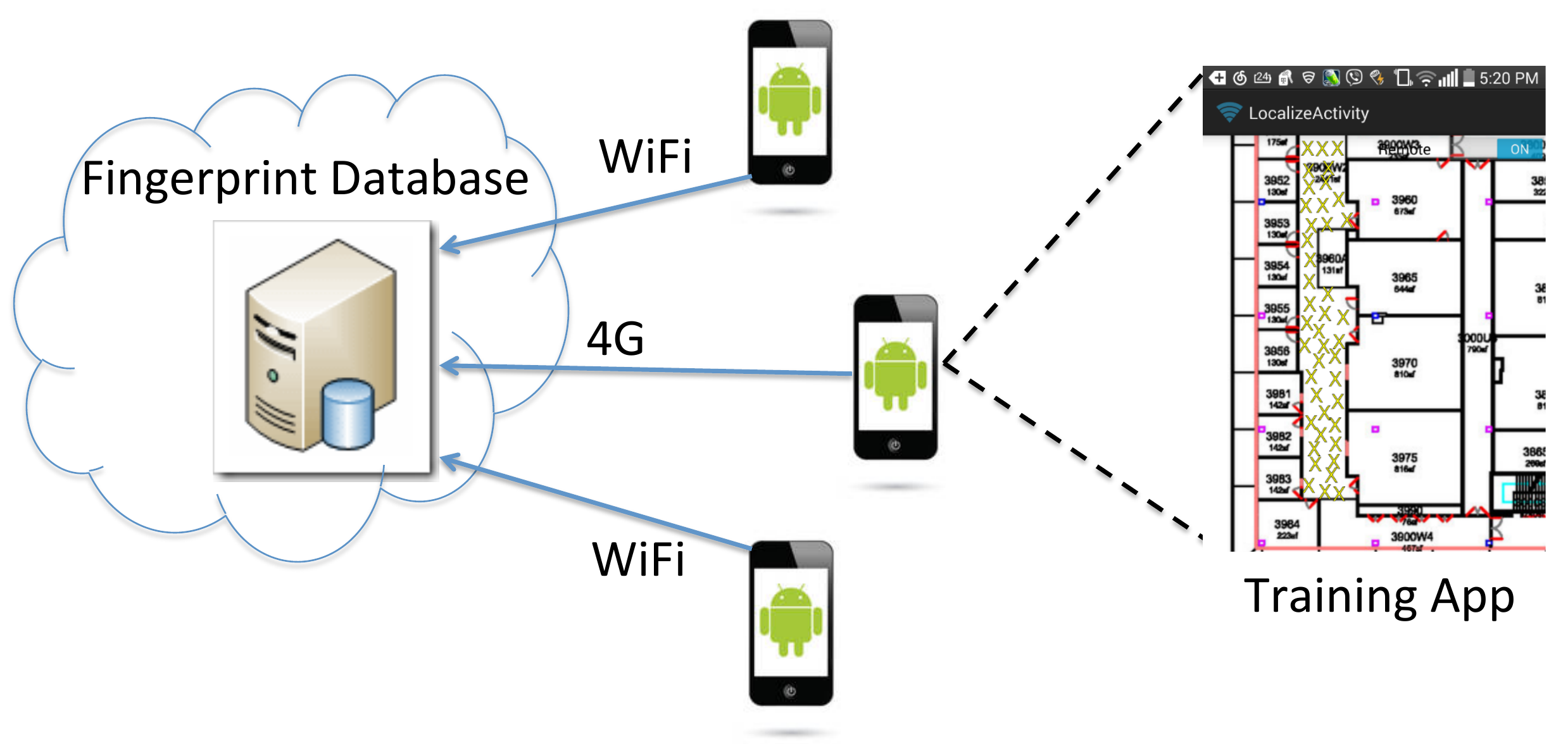} 
    \caption{Crowd sourcing-based training model \cite{ORIGINAL}.}
    \label{fig:training}
    \end{center}
\end{figure}

The idea behind indoor localization using fingerprints is based on the signals collected from APs for each user. Specifically, a user receives Wi-Fi signals from various APs at a specific location, which is referred to as a fingerprint, and these are stored in a database. At every location, this signal collection is done for a given area. The way these locations are picked is based on the expected accuracy of the localization. The higher the resolution of this process, the better the accuracy will be. To increase the accuracy, crowd sourcing can be used with some incentives. Once all location fingerprints are stored in a database, this information can now be used to determine the location of a user moving around. This phase of collecting fingerprints is called training as shown in Fig. \ref{fig:training}. 

After the training phase, the actual localization occurs when a user receives signals from an AP and then this signal is used to find out a similar signal in the database. Essentially, the closest signal which corresponds to a specific point in the database will be found. This process is achieved through a similarity distance computation. There may be different approaches to determine the closest distance. One simple approach is to find the single fingerprint while another approach might find $k$ closest fingerprints and take the centroid of these fingerprints \cite{ORIGINAL}.

\subsection{Homomorphic Encryption}
Homomorphic encryption enables computation on encrypted numbers so as to ensure privacy of the actual numbers. While most of the homomorphic systems are partial (i.e., they only support two arithmetic operations), recent years witnessed fully homomorphic systems that support all arithmetic operations \cite{van2010fully}. 
In this paper, we utilized the well-known Paillier \cite{paillier1999public} algorithm and DGK algorithm \cite{damgaard2007efficient} which are partially homomorphic systems. 
Specifically, these homomorphic encryption schemes share the following two traits which permit for computation on encrypted values. Let E($x$) and E($y$) be encrypted $x$ and $y$ respectively. Under homomorphic encryption E($x$)E($y$) = E($x$ + $y$). Also, they both support scalar multiplication: If we have E($x$) and plain-text y, we can compute E($xy$) by E$(x)^{y}$.

DGK has also the benefit of supporting a built-in comparison protocol for comparing encrypted numbers.
This protocol is one solution to the "socialist millionaire's" problem, where the objective is to generate a true inequality without decrypting the cipher-texts \cite{boudot2001fair}. Veugen \cite{veugen2018correction} describes Protocol 2 and Protocol 4 in his work to compare encrypted numbers. When using Protocol 4 to compare DGK values, it reliably computes $[[x > y]]$. 
The main benefit of Protocol 4 is that the constraints enforced on the plain-text space is much smaller, allowing for successful and secure comparisons with DGK encrypted values.

\subsection{Privacy in Fingerprint-based Localization}
One of the issues raised in fingerprinting is privacy. This occurs because the fingerprints collected from a user device are exposed to the server which stores all the fingerprints. Consequently, these fingerprints could be used to determine the locations of a specific user and track him/her during the day. To address this issue, our previous work in \cite{ORIGINAL} proposed employing a partial homomorphic system based on Paillier. This approach encrypts the fingerprints and when computing the closest distance at the server, it performs encrypted computation and eventually sends the encrypted results to the client to decrypt and compute the closest one. 

\section{DGK-based Fingerprint Localization}

In this section, we describe our proposed approach for fingerprint-based localization in details. 

\subsection{Overview}
Similar to the works in the literature such as \cite{ORIGINAL}, our approach utilizes homomorphic encryption when collecting data from users and doing computation at the database server. However, as previous works utilized homomorphic systems such as Paillier which cannot do comparison of encrypted numbers, the computed distance results at the server were being passed back to the client which can then decrypt these results and can find the closest distance. This requires communication of all of these values as well as decryption overhead at the client. In this paper, we propose to eliminate this overhead by utilizing a comparison protocol based on DGK and hence conduct comparison of encrypted distances at the server. The DGK comparison protocol helps us to determine the closest distance to user's fingerprint and then we communicate only that encrypted value to the user. We explain the details of this approach in the balance of this section. 

\subsection{Training Phase of the Proposed Approach}
We implemented a smart phone app that follows our proposed fingerprint-based approach. This smart phone app was started to be tested at Florida International University (FIU) Engineering Center but once the REU student left the program, it continued at his dormitory at Columbia University. 

The fingerprint-based application first requires fingerprints (training data) stored in a remote server that is accessible by the client's device via Wi-Fi or LTE.  In our case, the server uses a MySQL database to store the fingerprints and utilizes the JDBC driver to retrieve data from the database and obtain a floor map of Broadway dormitory third floor. The database contains a table with the following columns:
\begin{enumerate}
\item Map ID
\item Location on the map
\item MAC-address of an AP
\item RSS value associated with the AP
\item Device manufacturer, model, and software version
\item Date and time
\end{enumerate}

The Android app has a training activity which allows users to collect and store the fingerprints to the remote database as shown in Fig.~\ref{fig:training}. A user does this by going to the location where s/he wants to fingerprint and by tapping the screen on the matching location in the floor map. The app will collect the coordinates, APs detected and their corresponding RSS. Upon completion, a blue "X" appears informing the user the location is recorded in the finger database. 

We conducted our scans with a range of about 5 - 10 feet apart from each other to give a precise location. Whenever possible, we would conduct scans within rooms with closed doors as to give a more distinct signature. As a result, we obtained a fingerprint database as follows: 

$\mathcal{D} = \langle \left(x_i, y_i \right), V_i = \left\{RSS_{j}, AP_{j}\right\}^{N_{\rm AP}}_{j=1}\rangle^{N_{\rm F}}_{i=1}$ 

where $ \left(x_i, y_i \right)$ is the location of the fingerprint, $V_i$ is the fingerprint tuple comprised of $AP_i$ and its fingerprint RSS value.  $N_{\rm AP}$ is the total number of fingerprint reading scan result APs for a location and $N_{\rm F}$ is the total number of fingerprints in the database. 

\subsection{Fingerprint Database pre-processing}
Once the database has sufficient fingerprints, it needs to pre-process the training data to create look up tables. This is because not all the locations will see the same APs and some AP signal values might be empty in the database. 

Basically, if the AP is not found in the database, we use $v_c$ = -120 to automatically setup the missed constant determined by Parmengol et al. \cite{ORIGINAL}. 

When completed, the look up tables will list all $X,Y$ coordinates and their signal values for each AP's MAC address.  Upon completion of the pre-processing phase, the remote server will be capable of securing localizing a user. 

\subsection{Homomorphic Distance Computation}
The localization scheme works by computing the Euclidean distance of the user's ($AP, RSS$) pairs to each row on the look up table to generate $N_{\rm F}$ rows of distance and coordinate pairs. 

More specifically, a client first conducts a localization scan and obtains a list of ($AP$, $RSS$) pairs. The client then obtains the columns in the look up table.
The client organizes their scan results correspond with the columns of the server look up table. Then, the client sends the normalized ($AP$, $RSS$) pairs and the public key to the server. To ensure privacy is preserved, the RSS data is encrypted via homomorphic encryption. 

In summary, the algorithm computes the distance squared using homomorphic encryption. It exploits the fact that $d^2$ = $(x_{2} - x_{1})^2$ = $(x_{2}^{2}) + (-2x_{1})(x_{2}) + (x_{1}^{2})$. 
Here $x_{2}$ is the server's fingerprint RSS data and $x_1$ is the localization scan RSS. The terms $S_1$, $S_2$, and $S_3$ correspond to the terms in parenthesis respectively. The pseudo-code for this computation is given in Algorithm \ref{euclid}. 

\begin{algorithm}
\caption{MCA Distance Computation}
\label{euclid}
\begin{algorithmic}[1]
\State{$\textit{L} \gets \textit{gatherLookupTableData()}$}
\State{$\textit{scanAPs} \gets \textit{APs found in Localization Scan}$}
\For{$\text{each fprint = $((x_i,y_i), V_i)$ in L}$}
    \State{$\textit{$S_{i, 1}$}$ $\gets$ 0}
    \For {$\text{each AP in scanAPs}$}
        \If {$\textit{fprint contains AP}$}
            \State {$\textit{$S_{i, 1} += RSS_{i, j}^{2}$}$}
            \State {$\textit{\textit{$[[S_{i, 2}]]$} += $[[S_{2, j}]]^{RSS_{i, j}}$}$}
            \Else
                \State {$\textit{$S_{i, 1} += (v_c)^2$}$}
                \State {$\textit{\textit{$[[S_{i, 2}]]$} += $[[S_{2, j}]]^{v_c}$}$}
            \EndIf
    \EndFor
    \State {$\textit{$[[S_{i, 1}]]$} \gets \textit{encrypt($S_{i, 1}$)}$}
    \State {$\textit{$[[d_i]]$} \gets \textit{$[[S_{i,1}]]$ + $[[S_{i,2}]]$ + $[[S_3]]$}$}
    \State {$\textit{result} \gets \textit{$(x_i, y_i), d_i$}$}
    \State {$\textit{add result to resultList}$}
\EndFor
\State {$\text{return resultList}$}
\end{algorithmic}
\end{algorithm}

As seen, this algorithm returns a list (i.e., $resultList$) which basically includes all distance computations to client's fingerprint $x_1$. All the elements of this list is encrypted. We followed both Paillier and DGK to perform this operation as they both support partial homomorphism. 

\subsection{Server-based Minimum Distance Computation}
We propose to use DGK algorithm for processing the $resultList$ instead of passing it back to the client. Since DGK has the ability to compare encrypted numbers, the algorithm will compare all the elements and determine the top $k$ of these.  In our case, we followed an approach which computes the closest distance (i.e., top candidate) to a given fingerprint and reports it back to client. 

To obtain the $k$ smallest encrypted numbers, we used Bubble sort as our template as shown in Algorithm \ref{K-Min} and modified it to fit our needs. First, we modified line 2 to loop only $k$ times and we changed the condition to check if arr[i] $<$ arr[i + 1]. Upon running this protocol, the smallest values are located in ascending order ranging from arr[n - 1], arr[n - 2], ..., arr[length - 1 - (k - 1)]. The main benefit of using this system over sorting the encrypted array is that its overhead is O($nk$) rather than O($n^2$) to sort the whole encrypted array.  Considering that Protocol 4 can take about 0.3 - 0.5 seconds \cite{veugen2018correction} for one execution, it is imperative we minimize the number of comparisons.

\begin{algorithm}
\caption{Integrating Protocol 4 and Bubble Sort}
\label{K-Min}
\begin{algorithmic}[1]
\State{$n$ $\gets$ arr.length}
\For{i = 0 to k}
    \For{j = 0 to n - i - 1}
        \If{Protocol4(arr[j], arr[j + 1]) == 0}
            \State{Cipher-text temp = arr[j]}
            \State{arr[j] = arr[j+1]}
            \State{arr[j + 1] = temp}
        \EndIf
    \EndFor
\EndFor
\end{algorithmic}
\end{algorithm}

\begin{figure}[H]
\setlength{\abovecaptionskip}{0cm}
\setlength{\belowcaptionskip}{0cm}
\begin{center}
\includegraphics[width=1\linewidth, height=1.6in]{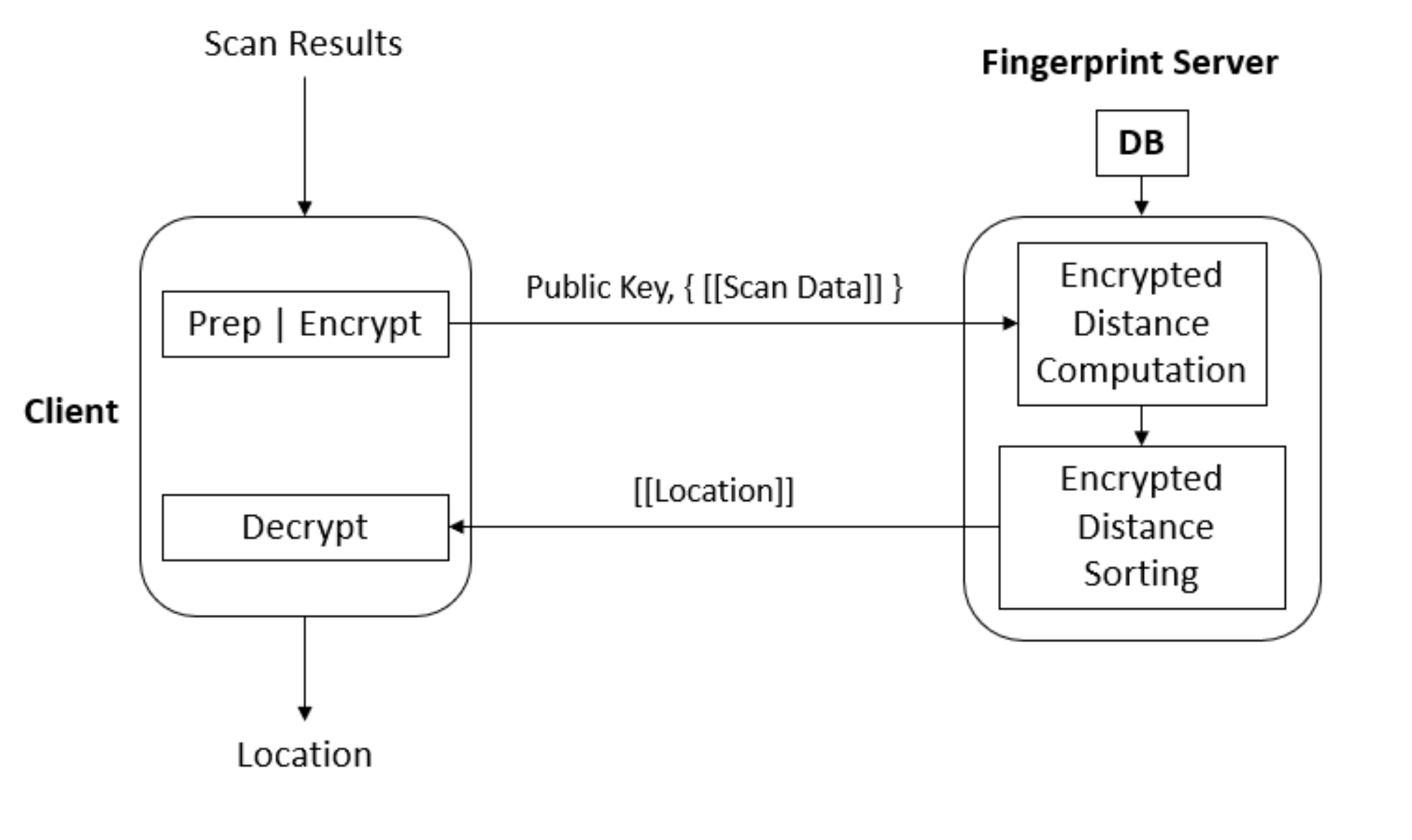}
\caption{Overall process for server-based fingerprinting localization. The use of brackets [[v]] on the value v denotes that the it is encrypted.}
\label{fig:model}
\end{center}
\end{figure}

Using Algorithm \ref{K-Min}, where $k$ = 1, we obtain the smallest encrypted distance. The server then returns the corresponding encrypted coordinates to the client. This overall process is summarized in Fig. \ref{fig:model}.

\section{Performance Evaluation}
In this section, we discuss the results of our experiments to test the performance of both our proposed approach. 

\subsection{Experiment Setup}
We have implemented Paillier and DGK in Java using the standard \textit{BigInteger} library, using the author's paper as reference. 
Also, we used a remote server which is a Desktop computer with an Intel 64-bit i7-4790K CPU running at 4 GHz with 24 GB of RAM. We have used the Samsung Galaxy 6S Edge which has a 8 core CPU running at 2.1GHz and 4 GB of RAM for both training and testing the performance of both localization systems. 

We created 19 fingerprints in our training data, in which we detected 265 APs. We had filtered 90\% of the APs so our column only consisted of 26 APs that were detected 8 - 13 times. 
We conducted 20 different trials for 2048 bits key lengths and we have kept track of how much battery was consumed upon completion.

\subsection{Performance Metrics and Baselines}
We considered the following metrics to access performance: 
\begin{itemize}
\item \textit{Execution Time}: This is the time for all computations and communication between the user and server. 
\item \textit{Battery Life}: This is defined as the energy consumption for the user device's battery. 

\end{itemize}

For comparison, we compared our server-side localization approach\footnote[1]{ https://github.com/AndrewQuijano/SSTREU2017} with Parmengol et al.'s \cite{ORIGINAL} client-side approach. For both approaches, we had the client send either DGK or Paillier encrypted distances. 
Therefore, our implementation of the comparison protocol can compare both Paillier and DGK cipher-text \footnote[2]{ https://github.com/AndrewQuijano/Homomorphic\_Encryption}. 
In both cases, we used 2048 bits of key sizes as 1024 bits is not considered secure enough anymore.

\subsection{Performance Results}

We first checked the execution time results. Based on our results in Table \ref{RESULTS}, we have observed an expected increase in time with the server-side approach. For instance, the client-side with Paillier took 11.13 seconds while it was 16.65 seconds in our DGK approach. However, we observed that 
DGK localization is much faster overall because DGK encryption and decryption operations are much faster compared to Paillier. 


\begin{table}[H]
\centering
\caption{Overhead of 20 localizations with 2048-bit keys}
\label{RESULTS}
\begin{tabular}{|c|c|c|c|c|}
\hline
Approach                    & Time (Paillier)   & Time (DGK) & Energy (Paillier/DGK)   \\ \hline
Client Side     & 11.13     & 6.71  & 4\% \\ \hline
Server Side     & 27.93     & 16.65 & 3\% \\ \hline
\end{tabular}
\end{table}

We then looked at the energy consumption for both cases. Note that using Paillier or DGK did not matter for client or server sides for energy consumption. Upon completion of our tests, the client system used about 4\% of the phone's battery and our system has used 3\% battery. This is because when analyzing Veugen's comparison protocol, the server is executing most of the computations, which was our objective. So if this localization computations are repeated many times, for instance, as part of a tracking app, then our server-based approach would be standing out for saving significant battery power. In other words, we can safely say that if we had repeated until the battery ran out, we would have completed more localizations using our model. This would be especially useful if the user highly valued their battery life over time to locate themselves such as a disaster scenario. In summary, DGK can be a viable option whether it is used in a client-based or server-based approach for privacy preserving indoor localization due to its speed and our server-side approach has energy saving features.
However, if we use DGK, we must ensure the distances doesn't exceed the plain-text space. This can be done by training more fingerprints and by only computing the distance to the closest fingerprints.

\section{Conclusion}
In this paper, we introduced a server-based fingerprint based localization scheme using DGK homomorphic encryption system. The goal was to push the comparison of encrypted distance values to the server so that the client can save energy and transmission time due to transmission of those values back to client in the existing approaches. We developed an Android app for indoor localization to test the proposed approach in student dormitories. Our results demonstrated that our system based on DGK can save battery life of the mobile device, while still preserving user privacy with a slight increase in execution time compared to the original method. However, we also found out that if implemented at the client-side, DGK can significantly outperform the existing approach with Paillier in terms of execution time. 

\section*{Acknowledgments}
Andrew Quijano was supported by the US NSF REU Site at FIU under the grant number REU CNS-1461119. The work is also supported in part by a grant from Cisco Silicon Valley Foundation. We would also like to thank Alice Niu for helping us with collecting experimental results and Dr. Samet Tonyali for his help in understanding the details of DGK.

\bibliographystyle{IEEEtran}
\bibliography{SSTREU2017}

\end{document}